\documentclass[aps,nofootinbib,commonaddress,preprint,showpacs]{revtex4}
\usepackage{amsmath}
\usepackage{amssymb}
\usepackage{graphicx}
\usepackage{epstopdf}
\usepackage{color}

\def\e{\begin{equation}}
\def\f{\end{equation}}

\begin{document}

\title{Guided modes in a spatially dispersive wire medium slab}
\author{Yu.~Tyshetskiy}
\email{yuriy.tyshetskiy@sydney.edu.au} \affiliation{School of Physics, The University of Sydney, NSW 2006, Australia}
\affiliation{Metamaterials Laboratory, National Research University of Information Technology, Mechanics, and Optics, St Petersburg 199034, Russia}
\author{S.V.~Vladimirov}
\affiliation{School of Physics, The University of Sydney, NSW 2006, Australia} 
\affiliation{Metamaterials Laboratory, National Research University of Information Technology, Mechanics, and Optics, St Petersburg 199034, Russia}
\author{A.E.~Ageyskiy}
\affiliation{Metamaterials Laboratory, National Research
University of Information Technology, Mechanics, and Optics, St
Petersburg 199034, Russia}
\author{I.~Iorsh}
\affiliation{Metamaterials Laboratory, National Research
University of Information Technology, Mechanics, and Optics, St
Petersburg 199034, Russia}
\author{A.~Orlov}
\affiliation{Metamaterials Laboratory, National Research
University of Information Technology, Mechanics, and Optics, St
Petersburg 199034, Russia}
\author{P.A.~Belov}
\affiliation{Metamaterials Laboratory, National Research
University of Information Technology, Mechanics, and Optics, St
Petersburg 199034, Russia}

\date{\today}
\received{}

\begin{abstract}
We study the guided modes in the wire medium slab taking into account both the nonlocality and losses in the structure. We show that due to the fact that the wire medium is an extremeley spatially dispersive metamaterial, the effect of nonlocality plays a critical role since it results in coupling between the otherwise orthogonal guided modes. We observe both the effects of strong and weak coupling, depending on the level of losses in the system.   

\end{abstract}
\pacs{78.20.Ci, 78.67.Pt, 41.20.Jb, 42.70.Qs, 78.70.Gq}

\maketitle


\newpage

\section{Introduction \label{sec:intro}}

Wire metamaterials have a number of unique properties~\cite{wirereview}, including the possibility of the subwavelength image transfer~\cite{Subwavelength1,Subwavelength2,Subwavelength3}, negative refraction~\cite{negrefr} and spontaneous emission time engineering~\cite{polar}. Localized  modes in the slabs of wire metamaterial have been studied previously in a number of papers~\cite{Theor,Belovmodes}, and it was shown that these modes are similar to the so-called spoof plasmons, a special class of surface modes, which  propagate along corrugated metal or semiconductor surfaces~\cite{spoof001,spoof01,spoof}. Furthermore, it has been shown that the excitation of the guided modes in the wire metamaterial slab is critical for the realization of the far-field superlensing~\cite{FinkPRL,Finkfirst}.

While the guided waves in the wire metamaterial slab have been studied previously~\cite{Finkwaveguides,FinkWCM}, there are no consistent studies of these modes that simultaneously account for the three distinctive features of these structures: the spatial dispersion of the dielectric permittivity (arising from the nonlocality of the wire medium's response to electromagnetic field), the presence of the wire host medium with dielectric permittivity different from that in vacuum, and the presence of the inevitable losses. At the same time, such an analysis is currently extremely demanded, since the realizations of the wire medium at the moment exist for a wide range of frequencies spanning from microwave to the optical range, and both spatial dispersion and essential losses are present in these samples. Thus, in order to correctly describe the effects being observed experimentally in the existing wire media samples, these features should be taken into account.

In this work, we present a consistent analysis of the properties of the guided waves in the wire metamaterial slabs, which accounts for all aforementioned effects. We present two approaches to obtaining the dispresion equations for the eigenmodes of the waveguiding metamaterial slabs, both yielding the same results. We then analyze the obtained band structure of the symmetric and antisymmetric eigenmodes of the waveguide. We start from recalling the results for the local case (when the dielectric permittivity of the metamaterial at a given position is assumed to be a function of that position only), and then compare with the obtained results for the nonlocal approach (when the dielectric permittivity at a given position is a function of both that position and its neighborhood). We find, in particular, that the nonlocal effects lead to a strong coupling between ``fast'' and ``slow'' eigenmodes of the waveguide, manifested by anti-crossings of their dispersion curves. We also study how the losses in the host media affect the dispersion and the coupling of the guided modes.

The paper is organized as follows: in section~\ref{sec:Setup} we present the detailed problem setup; in section~\ref{sec2} general impedance relations at the boundary of wire media slab are derived; section~\ref{sec:TM_sas} is dedicated to the derivation of the dispersion equation for the symmetric and antisymmetric  guided modes; band structures of the eigenmodes are presented in section \ref{sec:Band} and the profiles of the electric fields for different eigenmodes are presented in \ref{sec:Profiles}. In section~\ref{sec:losses} it is shown how the losses in the structure affect the dispersion properties of the eigenmodes, and the conclusions are presented in section~\ref{sec:concl}.

\section{Guided modes in a homogenized wire medium slab \label{Sec1}}
\subsection{Problem setup \label{sec:Setup}}
\begin{figure}
\includegraphics[width=3.0in]{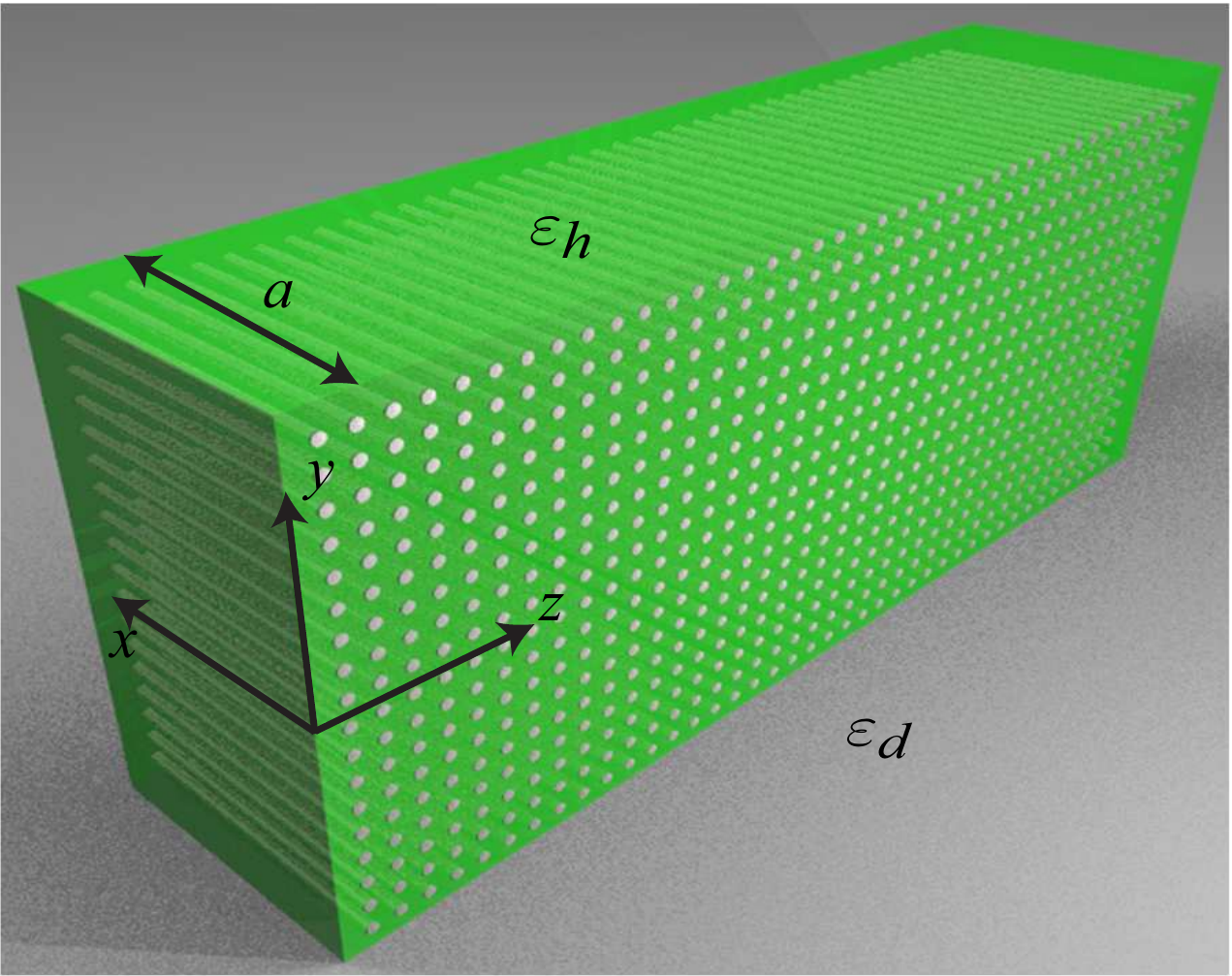}
\caption{\label{fig:setup} Waveguiding structure: a planar slab of
wire medium of thickness $a$, cladded on both sides by a uniform
dielectric with a constant permittivity $\varepsilon_d$. The wire
medium consists of parallel perfectly conducting wires embedded
into a uniform host medium with a dielectric permittivity
$\varepsilon_h$.}
\end{figure}

We consider a planar slab of wire medium (WM), composed of ideally
conducting parallel thin wires (oriented perpendicular to the slab
surfaces), which are embedded in a uniform host medium with a
constant dielectric permittivity $\varepsilon_h$. The slab has a
thickness $a$, occupying the region $0\leq x\leq a$, and is
cladded on both sides by a dielectric with a constant dielectric
permittivity $\varepsilon_d$, as shown in Fig.~\ref{fig:setup}.

We are interested in guided electromagnetic modes of such
structure, with wavelengths that are large compared to the largest
characteristic spatial scale of the wire medium (the latter is the
period of the wire array in WM). Under such conditions, the bulk
wire medium can be described as a \textit{uniaxial} spatially
dispersive (nonlocal) medium with an effective dielectric
permittivity tensor~\cite{Belov_etal_PRB_2003,Maslovski_PRB_2009}
(we use SI units throughout)
\begin{equation}
\boldsymbol{\varepsilon}_{\rm WM} =
\varepsilon_0\varepsilon_h\left(\varepsilon_{xx}(\omega,k_x)\
\mathbf{\hat{x}}\mathbf{\hat{x}} +
\mathbf{\hat{y}}\mathbf{\hat{y}} +
\mathbf{\hat{z}}\mathbf{\hat{z}}\right),  \label{eq:epsilon_WM}
\end{equation}
with
\begin{equation}
\varepsilon_{xx}(\omega,k_x) = 1 - \frac{\omega_{h0}^2}{\omega^2 -
c_h^2 k_x^2},   \label{eq:eps_xx}
\end{equation}
where $c_h=c/\sqrt{\varepsilon_h}$ is the speed of light in the host medium of the WM (we assume a non-magnetic host medium with $\mu_h=1$), $\omega_{h0}=\omega_p/\sqrt{\varepsilon_h}$, $\omega_p=c k_p$ is the ``plasma frequency'' of the WM with $k_p$ defined in Eq.~(10) of~\cite{Belov_etal_PRB_2003}, $k_x$ is the $\mathbf{k}$-vector component along the wires, $c$ is the speed of light in vacuum, and $\omega$ and $\mathbf{k}$ are the frequency and wavevector of a Fourier-transformed electromagnetic field associated with the electromagnetic modes of the unbounded WM.

We consider electromagnetic waves propagating along the WM slab;
choosing axis $z$ along the direction of propagation, we have,
without reducing generality of the problem, the following
structure of the electromagnetic waves propagating in the
considered structure:
\[
\{\mathbf{E,B}\} = \{\mathbf{E}(x),\mathbf{B}(x)\}
e^{i\left(\omega t - k_z z\right)}.
\]
We note that, since the WM dielectric tensor (\ref{eq:epsilon_WM})
is invariant with respect to rotations of coordinate frame around
axis $x$ (which is fixed by assuming that the wires are
perpendicular to the slab boundaries), its form remains the same
as (\ref{eq:epsilon_WM}) in the chosen coordinate frame with axis
$z$ directed along the direction of propagation of the guided
modes.

Writing the Maxwell's equations for $\mathbf{E,B}$ and applying
Fourier transforms with respect to $t$ and $z$ (but not $x$), we
obtain the following set of equations for the components of
Fourier-transformed fields $\mathbf{E}^{(\omega,k_z)}(x)$ and
$\mathbf{B}^{(\omega,k_z)}(x)$:
\begin{eqnarray}
i\omega B_x^{(\omega,k_z)} &=& -i k_z E_y^{(\omega,k_z)}, \nonumber \\
i\omega B_y^{(\omega,k_z)} &=& i k_z E_x^{(\omega,k_z)} -
\frac{\partial
E_z^{(\omega,k_z)}}{\partial x}, \nonumber \\
i\omega B_z^{(\omega,k_z)} &=& \frac{\partial
E_y^{(\omega,k_z)}}{\partial x},
\nonumber \\
-c^2\varepsilon_0 i k_z B_y^{(\omega,k_z)} &=& -i
\omega\varepsilon_0
E_x^{(\omega,k_z)} + j_x^{(\omega,k_z)}, \nonumber \\
c^2\varepsilon_0 i k_z B_x^{(\omega,k_z)} -
c^2\varepsilon_0\frac{\partial B_z^{(\omega,k_z)}}{\partial x} &=&
-i \omega\varepsilon_0 E_y^{(\omega,k_z)} +
j_y^{(\omega,k_z)}, \nonumber \\
c^2\varepsilon_0\frac{\partial B_y^{(\omega,k_z)}}{\partial x} &=&
-i \omega\varepsilon_0 E_z^{(\omega,k_z)} + j_z^{(\omega,k_z)},
\label{eq:Maxwell_wkz}
\end{eqnarray}
where $j_{x,y,z}^{(\omega,k_z)}$ are the components of the
Fourier-transformed (with respect to $t$ and $z$) effective
current $\mathbf{j}(t,\mathbf{r})$ in the corresponding medium
(either WM or a cladding dielectric in our case), defined by the
constitutive relations of the medium. Thus
Eqs~(\ref{eq:Maxwell_wkz}) with the corresponding effective
current densities are valid both in WM (at $0\leq x\leq a$) and in
the cladding dielectric ($x<0$ and $x>a$).

It can be easily seen that in a bulk WM with dielectric tensor
(\ref{eq:epsilon_WM}), as well as in a cladding dielectric with a
constant $\varepsilon_d$, one has $j_i^{(\omega,k_z)}\propto
E_i^{(\omega,k_z)}$, with index $i=x,y,z$. Therefore,
Eqs~(\ref{eq:Maxwell_wkz}) separate into two independent sets of
equations for $E_x,E_z,B_y$ (TM wave) and $E_y,B_x,B_z$ (TE wave).
Below we consider guided TM modes of the structure.

\subsection{Guided TM modes \label{sec2}}
For TM modes, the relevant equations from the complete set
(\ref{eq:Maxwell_wkz}) are
\begin{eqnarray}
i k_z E_x^{(\omega,k_z)} - \frac{\partial
E_z^{(\omega,k_z)}}{\partial x} -
i\omega B_y^{(\omega,k_z)} &=& 0, \nonumber \\
c^2\varepsilon_0 i k_z B_y^{(\omega,k_z)} -i \omega\varepsilon_0
E_x^{(\omega,k_z)} + j_x^{(\omega,k_z)} &=& 0, \nonumber \\
c^2\varepsilon_0\frac{\partial B_y^{(\omega,k_z)}}{\partial x} + i
\omega\varepsilon_0 E_z^{(\omega,k_z)} - j_z^{(\omega,k_z)} &=& 0.
\label{eq:TM_Maxwell_wkz}
\end{eqnarray}
The boundary conditions for Eqs~(\ref{eq:TM_Maxwell_wkz}) follow
from (i) continuity of tangential field components $E_z$ and
$B_y$, and (ii) arrest of the normal component $j_x$ of the
effective WM current density at the slab boundaries $x=0,a$.

We solve Eqs~(\ref{eq:TM_Maxwell_wkz}) in the wire medium ($0\leq
x\leq a$) using Fourier method; see Appendix~\ref{sec:app1} for details. As a result, we obtain the \textit{impedance relations} between
the tangential components of electric and magnetic fields at the WM slab boundaries $x=0,a$:
\begin{eqnarray}
E_z^{(\omega,k_z)}(0) = -i S_1 B_y^{(\omega,k_z)}(0) + i S_2
B_y^{(\omega,k_z)}(a), \label{eq:E_z(0)} \\
E_z^{(\omega,k_z)}(a) = -i S_2 B_y^{(\omega,k_z)}(0) + i S_1
B_y^{(\omega,k_z)}(a), \label{eq:E_z(a)}
\end{eqnarray}
with
\begin{eqnarray}
S_1 &=& \frac{2 c^2}{\omega\varepsilon_h
a}\left.\sum_{n=0}^\infty\right.^\prime{\frac{c^2k_z^2-\omega^2\varepsilon_h\varepsilon_{xx}(n)}{c^2k_z^2-\omega^2\varepsilon_h\varepsilon_{xx}(n)+c^2\alpha_n^2\varepsilon_{xx}(n)}},
\label{eq:S1_sum} \\
S_2 &=& \frac{2 c^2}{\omega\varepsilon_h
a}\left.\sum_{n=0}^\infty\right.^\prime{(-1)^n\frac{c^2k_z^2-\omega^2\varepsilon_h\varepsilon_{xx}(n)}{c^2k_z^2-\omega^2\varepsilon_h\varepsilon_{xx}(n)+c^2\alpha_n^2\varepsilon_{xx}(n)}},
\label{eq:S2_sum}
\end{eqnarray}
where $\varepsilon_{xx}(n)=\varepsilon_{xx}(\omega,k_x=\alpha_n)$,
and $\alpha_n=n\pi/a$. Below we consider the impedance relations
for nonlocal and local models of wire medium slabs.

\subsubsection{Nonlocal wire medium slab}
Substituting $\varepsilon_{xx}(n)$, corresponding to the spatially
dispersive model of WM with the \textit{nonlocal} response
(\ref{eq:eps_xx}), into (\ref{eq:S1_sum})--(\ref{eq:S2_sum}),
introducing dimensionless variables $\Omega=\omega/\omega_{h0}$,
$K_z=ck_z/\omega_{h0}$, $\tilde{\alpha}_n=n\pi/\tilde{a}$,
$\tilde{a}=(\omega_{h0}/c)a$, and carrying out the summation, we
obtain the impedance relations for a nonlocal WM slab model:
\begin{eqnarray}
S_1 = S_1^{\rm nl} &=&
\frac{c}{\varepsilon_h\Omega}\frac{1}{K_z^2+\varepsilon_h}\left[\frac{\varepsilon_h^{3/2}\Omega}{\tan\left(\tilde{a}\varepsilon_h^{1/2}\Omega\right)}
+
\frac{K_z^2\kappa_x}{\tanh\left(\tilde{a}\kappa_x\right)}\right],
\label{eq:S1_nl} \\
S_2 = S_2^{\rm nl} &=&
\frac{c}{\varepsilon_h\Omega}\frac{1}{K_z^2+\varepsilon_h}\left[\frac{\varepsilon_h^{3/2}\Omega}{\sin\left(\tilde{a}\varepsilon_h^{1/2}\Omega\right)}
+
\frac{K_z^2\kappa_x}{\sinh\left(\tilde{a}\kappa_x\right)}\right],
\label{eq:S2_nl}
\end{eqnarray}
where
$\kappa_x=\sqrt{K_z^2-\varepsilon_h\left(\Omega^2-1\right)}$.

\subsubsection{Local wire medium slab}
A local model of the WM slab was proposed
in~\cite{Luukkonen_etal_2008} as a quasi-static approximation of
the uniaxial wire medium considered here, with the nonlocal
$\varepsilon_{xx}$ of (\ref{eq:eps_xx}) replaced by its local
approximation
\begin{equation}
\varepsilon_{xx}^{\rm l} = 1 - \frac{\omega_{h0}^2}{\omega^2}.
\label{eq:eps_xx_local}
\end{equation}
With (\ref{eq:eps_xx_local}) substituted into
(\ref{eq:S1_sum})--(\ref{eq:S2_sum}), we obtain the impedance
relations for a local WM slab model (i.e., with the spatial
dispersion ignored):
\begin{eqnarray}
S_1 = S_1^{\rm l} &=& \frac{c}{\varepsilon_h\sqrt{\Omega^2-1}}
\frac{\kappa_x}{\tanh\left(\frac{\Omega}{\sqrt{\Omega^2-1}}\tilde{a}\kappa_x\right)},
\label{eq:S1_l} \\
S_2 = S_2^{\rm l} &=& \frac{c}{\varepsilon_h\sqrt{\Omega^2-1}}
\frac{\kappa_x}{\sinh\left(\frac{\Omega}{\sqrt{\Omega^2-1}}\tilde{a}\kappa_x\right)},
\label{eq:S2_l}
\end{eqnarray}
where again
$\kappa_x=\sqrt{K_z^2-\varepsilon_h\left(\Omega^2-1\right)}$.

The obtained impedance relations (\ref{eq:S1_nl})--(\ref{eq:S2_nl}) and
(\ref{eq:S1_l})--(\ref{eq:S2_l}) for nonlocal and local WM slabs,
respectively, together with the impedances of both semi-bounded
dielectrics cladding the slab, allow to find the dispersion
relations for guided modes of such slabs. Below we consider
symmetric and antisymmetric TM modes of both nonlocal and local WM
slabs.

\subsection{Dispersion equations for symmetric and antisymmetric TM modes \label{sec:TM_sas}}
There are two types of modes that can propagate in the WM slab: symmetric and antisymmetric.
In a symmetric TM wave the tangential electric field $E_z^{\omega,k_z}(x)$ is symmetric, and the
tangential magnetic field $B_y^{\omega,k_z}(x)$ is antisymmetric with respect to the WM slab mid-plane:
\[
E_z^{\omega,k_z}(0)=E_z^{\omega,k_z}(a);\ \ B_y^{\omega,k_z}(0) =
-B_y^{\omega,k_z}(a).
\]
Conversely, in an antisymmetric TM wave the tangential electric field $E_z^{\omega,k_z}(x)$ is antisymmetric,
and the tangential magnetic field $B_y^{\omega,k_z}(x)$ is symmetric with respect to the WM slab mid-plane:
\[
E_z^{\omega,k_z}(0)=-E_z^{\omega,k_z}(a);\ \ B_y^{\omega,k_z}(0) =
B_y^{\omega,k_z}(a).
\]

The WM slab impedance for the symmetric and antisymmetric TM modes is then
\begin{equation}
Z_{\rm WM}^{\rm s, as} =
\varepsilon_0c^2\frac{E_z^{\omega,k_z}(0)}{B_y^{\omega,k_z}(0)} =
-i\varepsilon_0c^2\left(S_1\pm S_2\right), \label{eq:Z_WM_sas}
\end{equation}
where the upper sign in the bracket corresponds to the symmetric, and the lower -- to the antisymmetric mode.

The impedance of the semi-bounded dielectric on either side of the
WM slab is~\cite{ABR_book}
\begin{equation}
Z_{\rm d} =
-i\frac{\varepsilon_0}{\varepsilon_d}\frac{c^3}{\Omega}\sqrt{K_z^2-\Omega^2\varepsilon_d}.
\end{equation}

Now, from the continuity of tangential field components $E_z$ and $B_y$
at the slab boundaries we obtain
\begin{equation}
Z_{\rm WM}^{\rm s,as} + Z_{\rm d} = 0,  \label{eq:Z_WM+Z_d_sas}
\end{equation}
which yields the dispersion equation for the symmetric and antisymmetric TM modes.
Below we obtain such dispersion equations for symmetric and antisymmetric TM modes
in nonlocal and local WM slabs.

\subsubsection{Nonlocal WM slab}
With (\ref{eq:S1_nl})--(\ref{eq:S2_nl}) substituted in
(\ref{eq:Z_WM_sas}), from (\ref{eq:Z_WM+Z_d_sas}) we obtain the
dispersion equations for symmetric and antisymmetric guided TM modes of the
\textit{nonlocal} WM slab model:
\begin{eqnarray}
\sqrt{K_z^2-\varepsilon_d\Omega^2} &+& \frac{\varepsilon_d}{\varepsilon_h}\frac{1}{K_z^2+\varepsilon_h} \left[\varepsilon_h^{3/2}\Omega\cot\left(\varepsilon_h^{1/2}\Omega\
\tilde{a}/2\right) \right.\nonumber \\
&+& \left. K_z^2\ \kappa_x\coth\left(\kappa_x\tilde{a}/2\right) \right] = 0,\text{  symmetric mode;} \label{eq:dispersion_TM_s_nl} \\
\sqrt{K_z^2-\varepsilon_d\Omega^2} &+& \frac{\varepsilon_d}{\varepsilon_h}\frac{1}{K_z^2+\varepsilon_h} \left[-\varepsilon_h^{3/2}\Omega\tan\left(\varepsilon_h^{1/2}\Omega\
\tilde{a}/2\right) \right.\nonumber \\ 
&+& \left. K_z^2\ \kappa_x\tanh\left(\kappa_x\tilde{a}/2\right) \right] = 0, \text{ antisymmetric mode.}
\label{eq:dispersion_TM_as_nl}
\end{eqnarray}

\subsubsection{Local WM slab}
With (\ref{eq:S1_l})--(\ref{eq:S2_l}) substituted in (\ref{eq:Z_WM_sas}), we obtain the dispersion equation for symmetric and antisymmetric guided TM modes of the \textit{local} WM slab model:
\begin{eqnarray}
\sqrt{K_z^2-\varepsilon_d\Omega^2} +
\frac{\varepsilon_d}{\varepsilon_h}\frac{\Omega}{\sqrt{\Omega^2-1}}
\kappa_x\coth\left[\frac{\kappa_x\tilde{a}}{2}\frac{\Omega}{\sqrt{\Omega^2-1}}\right]
&=& 0, \text{ symmetric mode;} \label{eq:dispersion_TM_s_l} \\
\sqrt{K_z^2-\varepsilon_d\Omega^2} +
\frac{\varepsilon_d}{\varepsilon_h}\frac{\Omega}{\sqrt{\Omega^2-1}}
\kappa_x\tanh\left[\frac{\kappa_x\tilde{a}}{2}\frac{\Omega}{\sqrt{\Omega^2-1}}\right]
&=& 0, \text{ antisymmetric mode.}  \label{eq:dispersion_TM_as_l}
\end{eqnarray}

We note that both symmetric and antisymmetric modes are guided by
the structure if the condition
\begin{equation}
\Omega<K_z/\sqrt{\varepsilon_d}   \label{eq:Omega_guided}
\end{equation}
is satisfied (i.e., if the wave propagating along the slab
undergoes a total internal reflection at the slab boundaries);
otherwise, the modes become leaky (i.e., their energy leaks away
from the slab in the form of radiation into the dielectric
cladding).

The obtained dispersion equations for symmetric and antisymmetric TM modes of the waveguide can also be derived by an alternative method, shown in Appendix~\ref{sec:Alter}. Both metods yield identical dispersion equations for the modes, which justifies their correctness. 

\section{Results and discussion}
\subsection{Band structure of WM slab with respect to guided TM modes \label{sec:Band}}
\begin{figure}[!h]
\includegraphics[width=0.9\textwidth]{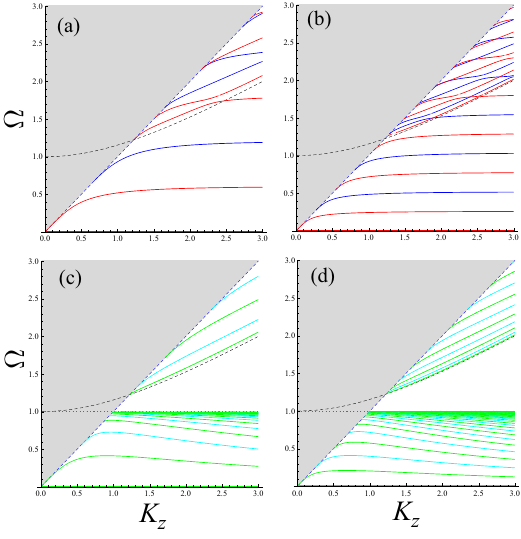}

\caption{\label{fig:TM_s+as_lnl} Band structure of symmetric and
antisymmetric guided TM modes in nonlocal ((a,b), blue curves for
symmetric, red curves for antisymmetric modes) and local ((c,d),
cyan curves for symmetric, green curves for antisymmetric modes)
WM slab models with $\varepsilon_h=3$, $\varepsilon_d=1$, for
different slab thicknesses $\tilde{a}=(\omega_{h0}/c)a$:
$\tilde{a}=3$ (a,c) and $\tilde{a}=7$ (b,d). The grey area
corresponds to the leaky modes, $\Omega>K_z/\sqrt{\varepsilon_d}$.
The dashed line $\Omega=\sqrt{1+K_z^2/\varepsilon_h}$ separates
``fast'' and ``slow'' guided modes, and the horizontal dotted line
marks the lower boundary of the mode band gap $\Omega=1$ in the
local WM slab model.}
\end{figure}

In this section we present the band diagrams of the wire media slab obtained within local (Eqs.~\eqref{eq:dispersion_TM_s_l},\eqref{eq:dispersion_TM_as_l}) and nonlocal (Eqs.~\eqref{eq:dispersion_TM_s_nl},\eqref{eq:dispersion_TM_as_nl}) models of wire medium, which are shown in Fig.\ref{fig:TM_s+as_lnl}.

It is immediately seen that the spatial dispersion of WM, due to the nonlocality of
$\varepsilon_{xx}$ in Eq.~(\ref{eq:eps_xx}), affects the band
structure of the WM slab in a qualitative way. Indeed, in the
local WM slab model characterized by (\ref{eq:eps_xx_local}) (with
the spatial dispersion ignored), there is a band gap, bounded by
the lines $\omega=\omega_{h0}$ and
$\omega=\sqrt{\omega_{h0}^2+c^2k_z^2/\varepsilon_h}$, in which the
guided modes do not exist at all. This band gap clearly separates the slow surface modes and the fast guided volume modes. Below this band gap, at $\omega<\omega_{h0}$, both symmetric and antisymmetric surface TM modes have regions of negative dispersion, corresponding to backward
waves with energy flowing in the direction opposite to their phase
velocities. In the nonlocal model of WM (\ref{eq:eps_xx}),
however, both these features of the band structure vanish, due to
the spatial dispersion: the former band gap now becomes filled
with the guided modes, and the regions of negative dispersion
disappear, so that both symmetric and antisymmetric guided modes
of a nonlocal WM slab have positive (normal) dispersion.

Moreover, an interesting band structure of guided modes appears in a nonlocal WM slab at frequencies
\begin{equation}
\sqrt{\omega_{h0}^2+\frac{c^2k_z^2}{\varepsilon_h}}<\omega<\frac{ck_z}{\sqrt{\varepsilon_d}}
\label{eq:Omega_fast}
\end{equation}
(for $\varepsilon_d<\varepsilon_h$), corresponding to the ``fast'' guided modes, as seen in Fig.~\ref{fig:TM_s+as_lnl}. Instead of series of almost parallel dispersion curves in the local WM slab model, in the nonlocal WM slab model the dispersion curves display the anticrossing behaviour. This effect is due to the coupling between different modes (the ``fast'' conventional waveguiding mode, and the ``slow'' surface mode) of the same parity, which is the effect conventionally observed in the coupled waveguide systems. The frequency splitting of the anti-crossing guided modes is proportional to the coupling strength, which in turn is proportional to the overlap integral of the two modes, which maximizes when the waveguide numbers of the two modes coincide. Near the anti-crossing points, the energy transfer between the coupled modes occurs, at the rate proportional to the coupling strength. Thus, it should be possible to excite both of the coupled modes by exciting only one mode of the pair with $K_z$ near their anti-crossing point. This can be particularly valuable for the excitation of the slow light modes by the free electromagnetic field (coupled to the fast waveguiding mode), which in its turn could be used in the optical information processing.

\subsection{Spatial structure of guided TM modes \label{sec:Profiles}}
The spatial structure of the fields
$E_x^{(\omega,k_z)}(x)$, $E_z^{(\omega,k_z)}(x)$,
$B_y^{(\omega,k_z)}(x)$ inside the slab is given by
Eqs~(\ref{eq:FT_x}), with $E_1(n)$, $E_3(n)$ and $B_2(n)$ obtained
from the linear system (\ref{eq:E1E3B2}) as
\begin{eqnarray}
E_1(n) &=& \frac{c^2}{\omega\varepsilon_h}\frac{k_z
\alpha_n}{c^2k_z^2-\omega^2\varepsilon_h\varepsilon_{xx}(n) +
c^2\alpha_n^2\varepsilon_{xx}(n)}\frac{2c^2}{a}\left[B_y^{(\omega,k_z)}(0)
-
(-1)^n B_y^{(\omega,k_z)}(a)\right], \\
E_3(n) &=& -i\frac{c^2}{\omega\varepsilon_h}\frac{c^2k_z^2 -
\omega^2\varepsilon_h\varepsilon_{xx}(n)}{c^2k_z^2-\omega^2\varepsilon_h\varepsilon_{xx}(n)
+
c^2\alpha_n^2\varepsilon_{xx}(n)}\frac{2}{a}\left[B_y^{(\omega,k_z)}(0)
-
(-1)^n B_y^{(\omega,k_z)}(a)\right], \\
B_2(n) &=& \frac{\alpha_n
\varepsilon_{xx}(n)}{c^2k_z^2-\omega^2\varepsilon_h\varepsilon_{xx}(n)
+
c^2\alpha_n^2\varepsilon_{xx}(n)}\frac{2c^2}{a}\left[B_y^{(\omega,k_z)}(0)
- (-1)^n B_y^{(\omega,k_z)}(a)\right],
\end{eqnarray}
in which the frequency $\omega$ is one of the solutions, for a
given $k_z$, of the relevant dispersion equation from those
obtained above,
Eqs~(\ref{eq:dispersion_TM_s_nl})--(\ref{eq:dispersion_TM_s_l})
and (\ref{eq:dispersion_TM_as_nl})--(\ref{eq:dispersion_TM_as_l}).

\begin{figure}
\includegraphics[width=0.9\textwidth]{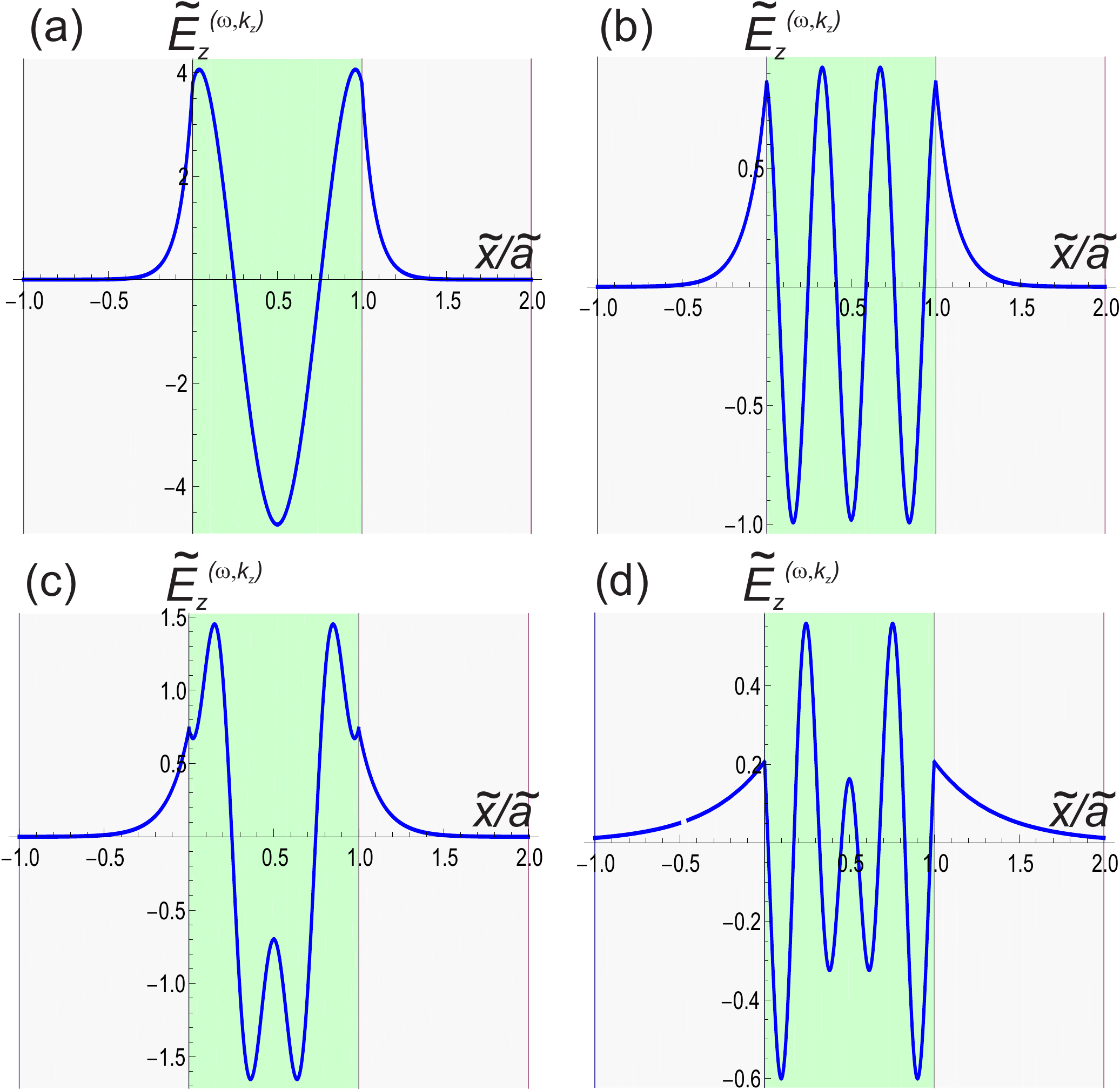}
\caption{\label{fig:spatial_sym} Spatial structure of $E_z^{(\omega,k_z)}(x)$ of symmetric guided TM modes with
$K_z=2.0$ in nonlocal WM slab model with $\varepsilon_h=3$,
$\varepsilon_d=1$, for slab thickness $\tilde{a}=7$. The upper
(a,b) and lower (c,d) rows show the spatial structure of the two
consecutive ``slow'' modes (with
$\Omega<\sqrt{1+K_z^2/\varepsilon_h}$) and the two consecutive
``fast'' modes (with $\Omega>\sqrt{1+K_z^2/\varepsilon_h}$),
respectively (see Fig.~\ref{fig:TM_s+as_lnl} for their band
structure at $K_z=2.0$).}
\end{figure}

The spatial structure of the $E_z^{(\omega,k_z)}(x)$ field of several consecutive symmetric modes in nonlocal WM slab is shown
in Fig.~\ref{fig:spatial_sym}.

\subsection{Effect of losses in the host medium \label{sec:losses}}
\begin{figure}[!h]
\includegraphics[width=0.95\textwidth]{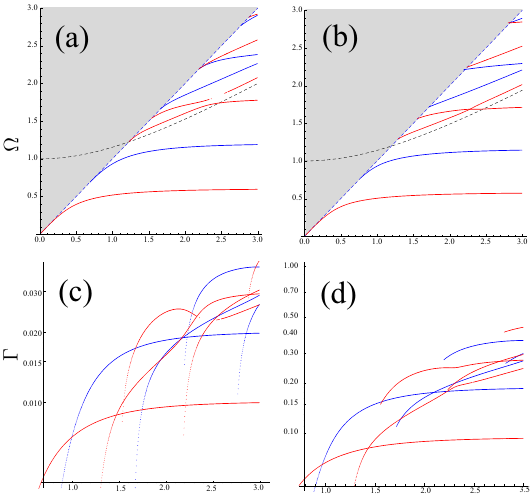}

\caption{\label{figloss} Band structure of the real (a,b) and negative imaginary (c,d) part of the eigenfrequencies for the case of low
$\mathrm{Im}(\varepsilon_h)=0.1$ (a,c) and high $\mathrm{Im}(\varepsilon_h)=1.0$ (b,d) losses, for nonlocal WM slab with $\tilde{a}=3$, $\varepsilon_d=1$, ${\rm Re}(\varepsilon_h)=3$.}
\end{figure}

Finally, we discuss the effect of losses present in the dielectric host media on the eigenmode dispersion. We introduce the losses by adding an imaginary part to the dielectric permittivity $\varepsilon_h$ of the host medium, and consider two different values of the imaginary part $\mathrm{Im}(\varepsilon_h)=0.1,1.0$ (with $\mathrm{Re}(\varepsilon_h)=3$), corresponding to weak and strong losses, respectively. The inclusion of the losses results in the emergence of negative imaginary part of the eigenfrequencies $\Gamma={\rm Im}(\Omega)<0$, which corresponds to the mode's damping rate. The band structure of $|\Gamma|$ is shown in Fig.~\ref{figloss}.

We can see that damping decrements of the modes increase and then saturate at their respective resonances. Moreover, we notice the transition from the strong mode coupling regime resulting in the anticrossing behaviour for the real parts of mode frequencies, to the weak-coupling regime resulting in the anticrossing of the imaginary parts of the frequencies as the losses are increased. This effect is also widely observed in the microcavity physics, when the transition from the strong to the weak coupling regime manifests itself in the transition from the anticrossing of the modes frequencies to the anticrossing of the modes decay rates.

\section{Conclusions \label{sec:concl}}
To conclude, we have studied the dispersion of the eigenmodes of the wire medium slab taking into account both the spatial dispersion and the losses in the structure. We have shown that the eigenmodes can be separated in two specific types: slow surface plasmon-polariton modes and fast conventional electromagnetic waveguide modes. We have also observed the strong pairwise coupling between slow and fast modes of the same parity resulting in the anticrossing of the dispersion curves. We stress that this effect arises only within the nonlocal approach, and thus has not been described previously. Moreover, we believe that the effect of the strong coupling between the slow  and fast modes can be particularly valuable for the excitation of the slow light modes by the free electromagnetic field, which in its turn could be used in the optical information processing.

\acknowledgments{This work was supported by the Australian
Research Council and by the Dynasty Foundation (Russia). The authors thank Yu. Kivshar and I.
Shadrivov for inspiring discussions. Yu.T. thanks Yu. Kivshar for
hospitality during his visit to ANU.}

\appendix
\section{Fourier method of solution for TM modes \label{sec:app1}}
Here we solve Eqs~(\ref{eq:TM_Maxwell_wkz}) for TM modes in the wire medium ($0\leq
x\leq a$) using Fourier method. For this, we continue the fields
and current densities, defined in the wire medium at $0\leq x\leq
a$, to the region $-a\leq x<0$ as~\cite{Kondratenko_Waveguides}
\begin{eqnarray}
E_x^{(\omega,k_z)}(-x) &=& -E_x^{(\omega,k_z)}(x)\ \text{(odd)} \nonumber \\
j_x^{(\omega,k_z)}(-x) &=& -j_x^{(\omega,k_z)}(x)\ \text{(odd)} \nonumber \\
E_z^{(\omega,k_z)}(-x) &=& E_z^{(\omega,k_z)}(x)\ \text{(even)}
\label{eq:flip}
\\
j_z^{(\omega,k_z)}(-x) &=& j_z^{(\omega,k_z)}(x)\ \text{(even)} \nonumber \\
B_y^{(\omega,k_z)}(-x) &=& -B_y^{(\omega,k_z)}(x)\ \text{(odd)}
\nonumber
\end{eqnarray}
and then continue these fields and current densities, now defined
at $-a\leq x\leq a$, periodically to the entire $x$ axis, with a
period of $2a$: $E_x^{(\omega,k_z)}(x+2a)=E_x^{(\omega,k_z)}(x)$,
etc. Thus continued fields and current densities are defined for
all $x$, and coincide with the physical fields and current
densities inside the wire medium slab. This mathematical trick
allows us to solve Eqs~(\ref{eq:TM_Maxwell_wkz}) for fields inside
the WM slab by seeking the solutions of
Eqs~(\ref{eq:TM_Maxwell_wkz}), for thus continued fields and
current densities, in the form of Fourier series
\begin{eqnarray}
E_x^{(\omega,k_z)}(x) &=& \sum_{n=0}^{\infty}{E_1(n) \sin(\alpha_n
x)},
\nonumber \\
j_x^{(\omega,k_z)}(x) &=& \sum_{n=0}^{\infty}{j_1(n) \sin(\alpha_n
x)},
\nonumber \\
E_z^{(\omega,k_z)}(x) &=&
\left.\sum_{n=0}^{\infty}\right.^\prime{E_3(n)
\cos(\alpha_n x)}, \label{eq:FT_x} \\
j_z^{(\omega,k_z)}(x) &=&
\left.\sum_{n=0}^{\infty}\right.^\prime{j_3(n)
\cos(\alpha_n x)}, \nonumber \\
B_y^{(\omega,k_z)}(x) &=& \sum_{n=0}^{\infty}{B_2(n) \sin(\alpha_n
x)}, \nonumber
\end{eqnarray}
where $\alpha_n=n\pi/a$, and $\sum^\prime$ implies that the $n=0$
term of the sum should be multiplied by $1/2$. Note that in
writing the above Fourier series, the symmetries of the continued
fields and current densities, introduced by Eqs~(\ref{eq:flip}),
have been taken into account. (Note that, in order to find the
physical fields outside the slab, one needs to solve a separate
problem for fields at $x<0$ and $x>a$, with $j_x=j_z=0$, and then
match the solutions for the obtained fields inside and outside the
slab, using continuity of tangential components of $\mathbf{E,B}$
at the slab boundaries $x=0,\ a$. This yields the dispersion
equation for the modes of such system; see Sec~\ref{sec:TM_sas}.)

Under the condition of zero normal component of the effective WM
current density at the WM slab boundaries, $j_x(x=0)=j_x(x=a)=0$,
the constitutive relations for the WM slab are the same as those
for the unbounded WM, and the Fourier coefficients of the current
densities $j_x$ and $j_z$ are obtained
as~\cite{Kondratenko_Waveguides}
\begin{equation}
j_i(n) = \sum_{k=1}^{3}\sigma_{ik}(n) E_k(n), \label{eq:ji(n)}
\end{equation}
where $\sigma_{ik}(n)$ is the conductivity tensor of the bulk WM,
in which $k_x=\alpha_n$:
\begin{equation}
\sigma_{ik}(n) = i\omega\varepsilon_0\left[\delta_{ik} -
\varepsilon_{ik}(n)\right], \label{eq:sigma_ik(n)}
\end{equation}
where $\varepsilon_{ik}(n)$ is given by (\ref{eq:epsilon_WM}) with
$k_x=\alpha_n=n\pi/a$.

From Eqs~(\ref{eq:TM_Maxwell_wkz}), using (\ref{eq:ji(n)}) and
(\ref{eq:sigma_ik(n)}), we obtain, taking into account that the
$B_y^{(\omega,k_z)}(x)$ field is discontinuous at the boundaries
$x=0,a$:
\begin{eqnarray}
i k_z E_1(n) + \alpha_n E_3(n) - i\omega B_2(n) &=& 0, \nonumber \\
-i\omega \varepsilon_h\varepsilon_{xx}(n) E_1(n) + c^2 ik_z B_2(n)
&=& 0,
\nonumber \\
i\omega\varepsilon_h E_3(n) + c^2\alpha_n B_2(n) &=&
\frac{2c^2}{a}\left[B_y^{(\omega,k_z)}(0) - (-1)^n
B_y^{(\omega,k_z)}(a)\right].
 \label{eq:E1E3B2}
\end{eqnarray}
From this linear system, we obtain $E_3(n)$ as
\begin{equation}
E_3(n) =
-i\frac{c^2}{\omega\varepsilon_h}\frac{c^2k_z^2-\omega^2\varepsilon_h\varepsilon_{xx}(n)}{c^2k_z^2-\omega^2\varepsilon_h\varepsilon_{xx}(n)+c^2\alpha_n^2\varepsilon_{xx}(n)}\frac{2}{a}\left[B_y^{(\omega,k_z)}(0)
- (-1)^n B_y^{(\omega,k_z)}(a)\right]. \label{eq:E_3(n)}
\end{equation}
Finally, substituting $E_3(n)$ into $E_z^{(\omega,k_z)}(x)=\sum^\prime{E_3(n) \cos(\alpha_n x)}$, and
taking $x=0,a$, we obtain the \textit{impedance relations} between the tangential components of electric and magnetic fields at the
WM slab boundaries, shown in Eqs~(\ref{eq:E_z(0)})--(\ref{eq:E_z(a)}).

\section{Alternative method \label{sec:Alter}}
In this section we present an alternative way to obtain the
dispersion of the eigenmodes in the wire medium slab based on the
additional boundary conditions technique~\cite{MGS}. Within the
local homogenization model for the case of perfectly conducting
wires the principal components of the dielectric permittivity
tensor become $[\infty,\varepsilon_h,\varepsilon_h]$. In this case
only a TEM polarized mode (a kind of modes that is useful for
transmission line mode's polarization, electric and magnetic
components of such type of modes are perpendicular to the
direction of propagation) can be excited in the wire media slab.
The eigenmode dispersion equation can then be recovered by
applying conventional continuity boundary conditions for electric
and magnetic fields at both interfaces of the slab. However, the
situation becomes more complicated if we account for
nonlocality~\cite{Maslovski_PRB_2009}, i.e. spatial dispersion of
the dielectric permittivity. In this case, wire media slab
supports propagation of both TEM and TM polarized waves. It is
evident then that we need an additional boundary condition in
order to obtain the eigenmode dispersion equation. The additional
boundary condition states that the current at the ends of the
wires is exactly zero and can be rewritten in the following form:
\begin{align}
\left[\frac{\partial^2H_z}{\partial
x^2}+\varepsilon_{h}\left(\frac{\omega}{c}\right)^2H_{z}\right]_{x=0,a}=0,  \label{eq:j=0}
\end{align}
where square brackets denote the difference at the corresponding interface: e.g.,
$[f(x)]_{x=0}=f(x=0^+)-f(x=0^-)$. Now when we have the three boundary
conditions (continuity of tangential components of electric and magnetic fields, and Eq.~(\ref{eq:j=0})) at each interface ($x=0,a$), we write down the fields inside the
wire medium slab. We would like to obtain both the eigenmode
dispersion equation and the transmission and reflection
coefficients for the wire medium slab and thus we consider the
case when the electromagnetic field $(\mathbf{E}_{\rm
inc},\mathbf{H}_{\rm inc})$ is incident on the wire medium slab.
The electromagnetic fields in the three regions can be written in
the form~\cite{AE,AE2}:
 \e \frac{H(x)}{H_{\rm inc}} =\left\{
\begin{array}{lcl}
e^{ik_xx} + R e^{-ik_xx},\quad x<0 \\[2mm]
\begin{array}{lcl}
A_-^{\rm TM}e^{-\frac{k_{p}\kappa_x}{\sqrt{\varepsilon_{h}}}(x-a/2)}\\
+ A_+^{\rm TM}e^{+\frac{k_{p}\kappa_x}{\sqrt{\varepsilon_{h}}}(x-a/2)}\\
+A_-^{\rm TEM}e^{ik_{p}\Omega(x-a/2)}\\
+ A_+^{\rm TEM}e^{-ik_{p}\Omega(x-a/2)},\\
\end{array} \quad 0\le x\le a \\[9mm]
T e^{ik_x(x-a)},\quad x>a, \vphantom{\frac{A}{A}}\\
\end{array}\right.
\label{eq:H} \f where $R$ and $T$ are unknown reflection and
transmission coefficients. $A_\pm^{\rm TM}$ and $A_\pm^{\rm TEM}$
are unknown amplitudes of TM and TEM waves that correspond to
forward and backward propagating waves along the $x$ axis from
Fig.~\ref{fig:setup}. We normalized this system to the magnetic
field of the incident wave $H_{\rm inc}$. If we then apply three
boundary conditions at each interface we get the linear system for
the amplitudes of the fields:
\begin{widetext}

\begin{eqnarray} \left(
  \begin{array}{cccccc}
  -1 & e^{\frac{k_{p}\kappa_x}{\sqrt{\varepsilon_{h}}}a/2} & e^{-\frac{k_{p}\kappa_x}{\sqrt{\varepsilon_{h}}}a/2} & e^{-ik_{p}a\Omega/2} & e^{ik_{p}a\Omega/2} & 0 \\
  ik_x & -\frac{k_{p}\kappa_x}{\varepsilon_{h}^{3/2}}e^{\frac{k_{p}\kappa_x}{\sqrt{\varepsilon_{h}}}a/2} & \frac{k_{p}\kappa_x}{\varepsilon_{h}^{3/2}}e^{-\frac{k_{p}\kappa_x}{\sqrt{\varepsilon_{h}}}a/2} & \frac{ik_{p}\Omega}{\varepsilon_{h}}e^{-ik_{p}a\Omega/2} & -\frac{ik_{p}\Omega}{\varepsilon_{h}}e^{ik_{p}a\Omega/2} & 0 \\
  k_x^2-\left(\frac{\omega}{c}\right)^2 &qe^{\frac{k_{p}\kappa_x}{\sqrt{\varepsilon_{h}}}a/2} & qe^{-\frac{k_{p}\kappa_x}{\sqrt{\varepsilon_{h}}}a/2} & 0 & 0 & 0 \\
  0 & e^{-\frac{k_{p}\kappa_x}{\sqrt{\varepsilon_{h}}}a/2} & e^{\frac{k_{p}\kappa_x}{\sqrt{\varepsilon_{h}}}a/2} & e^{ik_{p}a\Omega/2} & e^{-ik_{p}a\Omega/2} & -1\\
  0 & -\frac{k_{p}\kappa_x}{\varepsilon_{h}^{3/2}}e^{-\frac{k_{p}\kappa_x}{\sqrt{\varepsilon_{h}}}a/2} & \frac{k_{p}\kappa_x}{\varepsilon_{h}^{3/2}}e^{\frac{k_{p}\kappa_x}{\sqrt{\varepsilon_{h}}}a/2} &  \frac{ik_{p}\Omega}{\varepsilon_{h}}e^{ik_{p}a\Omega/2} & -\frac{ik_{p}\Omega}{\varepsilon_{h}}e^{-ik_{p}a\Omega/2} & -ik_x\\
  0 & qe^{-\frac{k_{p}\kappa_x}{\sqrt{\varepsilon_{h}}}a/2} & qe^{\frac{k_{p}\kappa_x}{\sqrt{\varepsilon_{h}}}a/2} & 0 & 0 & k_x^2-\left(\frac{\omega}{c}\right)^2\\
\end{array}
\right)\nonumber
\\
 \times\left(
  \begin{array}{c}
    R\\
    A_-^{\rm TM}\\
    A_+^{\rm TM}\\
    A_-^{\rm TEM}\\
    A_+^{\rm TEM}\\
    T\\
  \end{array}
\right) = \left(
  \begin{array}{c}
     1\\
     ik_x\\
     -k_x^2+\left(\frac{\omega}{c}\right)^2\\
     0\\
     0\\
     0\\
  \end{array}
\right), \qquad \label{eq:syst}\end{eqnarray}
\end{widetext}
where
$q=\left(k_{p}\kappa_x/\sqrt{\varepsilon_{h}}\right)^2+\varepsilon_{h}\left(\omega/c\right)^2$.
Solving the system and changing to dimensionless variables
introduced above in Sec~\ref{sec2}, we get the transmission and
reflection coefficients as well as amplitudes of all the modes in
the structure:

\begin{widetext}
\begin{eqnarray}
T= \left[1+\frac{1}{\varepsilon_h\left(K_z^2+\varepsilon_h\right)\sqrt{K_z^2-\Omega^2}}\left(-\varepsilon_h^{3/2}\Omega\tan\left(\varepsilon_h^{1/2}\Omega\tilde{a}/2\right)+K_z^2\kappa_x\tanh\left(\kappa_x\tilde{a}/2\right)\right)\right]^{-1}\nonumber\\
-\left[1+\frac{1}{\varepsilon_h\left(K_z^2+\varepsilon_h\right)\sqrt{K_z^2-\Omega^2}}\left(\varepsilon_h^{3/2}\Omega\cot\left(\varepsilon_h^{1/2}\Omega\tilde{a}/2\right) + K_z^2\kappa_x\coth\left(\kappa_x\tilde{a}/2\right)\right)\right]^{-1},\nonumber\\
 \label{eq:T}
\end{eqnarray}
\begin{eqnarray}
R=
\left[1+\frac{1}{\varepsilon_h\left(K_z^2+\varepsilon_h\right)\sqrt{K_z^2-\Omega^2}}\left(-\varepsilon_h^{3/2}\Omega\tan\left(\varepsilon_h^{1/2}\Omega\tilde{a}/2\right)+K_z^2\kappa_x\tanh\left(\kappa_x\tilde{a}/2\right)\right)\right]^{-1}\nonumber\\
+\left[1+\frac{1}{\varepsilon_h\left(K_z^2+\varepsilon_h\right)\sqrt{K_z^2-\Omega^2}}\left(\varepsilon_h^{3/2}\Omega\cot\left(\varepsilon_h^{1/2}\Omega\tilde{a}/2\right) + K_z^2\kappa_x\coth\left(\kappa_x\tilde{a}/2\right)\right)\right]^{-1}-1,\nonumber\\
 \label{eq:R}
\end{eqnarray}
\begin{eqnarray}
A_-^{TM}=
\frac{1}{2\left(1+\frac{\varepsilon_{h}}{K_z^2}\right)}\left(\frac{\mbox{sech}\left(\kappa_x\tilde{a}/2\right)}{1+\frac{1}{\varepsilon_h\left(K_z^2+\varepsilon_h\right)\sqrt{K_z^2-\Omega^2}}\left(-\varepsilon_h^{3/2}\Omega\tan\left(\varepsilon_h^{1/2}\Omega\tilde{a}/2\right)+K_z^2\kappa_x\tanh\left(\kappa_x\tilde{a}/2\right)\right)}\right.\nonumber\\
\left.+\frac{\mbox{csch}\left(\kappa_x\tilde{a}/2\right)}{1+\frac{1}{\varepsilon_h\left(K_z^2+\varepsilon_h\right)\sqrt{K_z^2-\Omega^2}}\left(\varepsilon_h^{3/2}\Omega\cot\left(\varepsilon_h^{1/2}\Omega\tilde{a}/2\right) + K_z^2\kappa_x\coth\left(\kappa_x\tilde{a}/2\right)\right)}\right),\nonumber\\
 \label{eq:ATMM}
\end{eqnarray}
\begin{eqnarray}
A_+^{TM}=
\frac{1}{2\left(1+\frac{\varepsilon_{h}}{K_z^2}\right)}\left(\frac{\mbox{sech}\left(\kappa_x\tilde{a}/2\right)}{1+\frac{1}{\varepsilon_h\left(K_z^2+\varepsilon_h\right)\sqrt{K_z^2-\Omega^2}}\left(-\varepsilon_h^{3/2}\Omega\tan\left(\varepsilon_h^{1/2}\Omega\tilde{a}/2\right)+K_z^2\kappa_x\tanh\left(\kappa_x\tilde{a}/2\right)\right)}\right.\nonumber\\
\left.-\frac{\mbox{csch}\left(\kappa_x\tilde{a}/2\right)}{1+\frac{1}{\varepsilon_h\left(K_z^2+\varepsilon_h\right)\sqrt{K_z^2-\Omega^2}}\left(\varepsilon_h^{3/2}\Omega\cot\left(\varepsilon_h^{1/2}\Omega\tilde{a}/2\right) + K_z^2\kappa_x\coth\left(\kappa_x\tilde{a}/2\right)\right)}\right),\nonumber\\
 \label{eq:ATMP}
\end{eqnarray}
\begin{eqnarray}
A_-^{TEM}=
\frac{1}{2\left(1+\frac{K_z^2}{\varepsilon_{h}}\right)}\left(\frac{\mbox{sec}\left(\varepsilon_h^{1/2}\Omega\tilde{a}/2\right)}{1+\frac{1}{\varepsilon_h\left(K_z^2+\varepsilon_h\right)\sqrt{K_z^2-\Omega^2}}\left(-\varepsilon_h^{3/2}\Omega\tan\left(\varepsilon_h^{1/2}\Omega\tilde{a}/2\right)+K_z^2\kappa_x\tanh\left(\kappa_x\tilde{a}/2\right)\right)}\right.\nonumber\\
\left.+\frac{i\mbox{cosec}\left(\varepsilon_h^{1/2}\Omega\tilde{a}/2\right)}{1+\frac{1}{\varepsilon_h\left(K_z^2+\varepsilon_h\right)\sqrt{K_z^2-\Omega^2}}\left(\varepsilon_h^{3/2}\Omega\cot\left(\varepsilon_h^{1/2}\Omega\tilde{a}/2\right) + K_z^2\kappa_x\coth\left(\kappa_x\tilde{a}/2\right)\right)}\right),\nonumber\\
 \label{eq:ATEMM}
\end{eqnarray}
\begin{eqnarray}
A_+^{TEM}=
\frac{1}{2\left(1+\frac{K_z^2}{\varepsilon_{h}}\right)}\left(\frac{\mbox{sec}\left(\varepsilon_h^{1/2}\Omega\tilde{a}/2\right)}{1+\frac{1}{\varepsilon_h\left(K_z^2+\varepsilon_h\right)\sqrt{K_z^2-\Omega^2}}\left(-\varepsilon_h^{3/2}\Omega\tan\left(\varepsilon_h^{1/2}\Omega\tilde{a}/2\right)+K_z^2\kappa_x\tanh\left(\kappa_x\tilde{a}/2\right)\right)}\right.\nonumber\\
\left.-\frac{i\mbox{cosec}\left(\varepsilon_h^{1/2}\Omega\tilde{a}/2\right)}{1+\frac{1}{\varepsilon_h\left(K_z^2+\varepsilon_h\right)\sqrt{K_z^2-\Omega^2}}\left(\varepsilon_h^{3/2}\Omega\cot\left(\varepsilon_h^{1/2}\Omega\tilde{a}/2\right) + K_z^2\kappa_x\coth\left(\kappa_x\tilde{a}/2\right)\right)}\right),\nonumber\\
 \label{eq:ATEMP}
\end{eqnarray}
So we get dispersion relations explicitly from the denominator of
expressions (\ref{eq:T})--(\ref{eq:ATEMP}):
\begin{eqnarray}
1+\frac{1}{\varepsilon_h\left(K_z^2+\varepsilon_h\right)\sqrt{K_z^2-\Omega^2}}\left(-\varepsilon_h^{3/2}\Omega\tan\left(\varepsilon_h^{1/2}\Omega\tilde{a}/2\right)+K_z^2\kappa_x\tanh\left(\kappa_x\tilde{a}/2\right)\right)=0,
 \label{eq:E1}
\end{eqnarray}
\begin{eqnarray}
1+\frac{1}{\varepsilon_h\left(K_z^2+\varepsilon_h\right)\sqrt{K_z^2-\Omega^2}}\left(\varepsilon_h^{3/2}\Omega\cot\left(\varepsilon_h^{1/2}\Omega\tilde{a}/2\right)
+ K_z^2\kappa_x\coth\left(\kappa_x\tilde{a}/2\right)\right)=0.
 \label{eq:E2}
\end{eqnarray}
\end{widetext}
These formulae fully coincide with
(\ref{eq:dispersion_TM_s_nl},\ref{eq:dispersion_TM_as_nl}). This
fact confirms the results of both methods.

\end{document}